\documentclass{article}

\usepackage[nonatbib,preprint]{tackling_climate_workshop_style}
\usepackage[utf8]{inputenc} 
\usepackage[T1]{fontenc}    
\usepackage{hyperref}       
\usepackage{url}            
\usepackage{booktabs}       
\usepackage{amsfonts}       
\usepackage{nicefrac}       
\usepackage{microtype}      
\usepackage{graphicx}
\usepackage{amsmath,amssymb,amsfonts,dsfont}
\usepackage{mathtools} 
\usepackage{xcolor}
\usepackage{cite}
\usepackage{gensymb}
\DeclareMathOperator*{\argmin}{argmin} 
\title{DL-Corrector-Remapper: A grid-free bias-correction deep learning methodology for data-driven high-resolution global weather forecasting}

\author{%
  Tao Ge\thanks{Worked performed during an internship at NVIDIA}\\
  Washington University in St. Louis\\
  St. Louis, MO. 63130 \\
  \texttt{getao@wustl.edu} \\
   \And
   Jaideep Pathak \\
  NVIDIA\\
  Santa Clara, CA. 95050 \\
  \texttt{jpathak@nvidia.com} \\
   \AND
   Akshay Subramaniam \\
  NVIDIA\\
  Santa Clara, CA. 95050 \\
  \texttt{asubramaniam@nvidia.com} \\
   \And
   Karthik Kashinath \\
  NVIDIA\\
  Santa Clara, CA. 95050 \\
  \texttt{kkashinath@nvidia.com} \\
}

\begin{document}

\maketitle

\begin{abstract}
Data-driven models, such as FourCastNet (FCN), have shown exemplary performance in high-resolution global weather forecasting. This performance, however, is based on supervision on mesh-gridded weather data without the utilization of raw climate observational data, the gold standard ground truth. In this work we develop a methodology to correct, remap, and fine-tune gridded uniform forecasts of FCN so it can be directly compared against observational ground truth, which is sparse and non-uniform in space and time. This is akin to bias-correction and post-processing of numerical weather prediction (NWP), a routine operation at meteorological and weather forecasting centers across the globe. The Adaptive Fourier Neural Operator (AFNO) architecture is used as the backbone to learn continuous representations of the atmosphere. The spatially and temporally non-uniform output is evaluated by the non-uniform discrete inverse Fourier transform (NUIDFT) given the output query locations. We call this network the Deep-Learning-Corrector-Remapper (DLCR). The improvement in DLCR’s performance against the gold standard ground truth over the baseline’s performance shows its potential to correct, remap, and fine-tune the mesh-gridded forecasts under the supervision of observations. 
\end{abstract}

\section{Introduction}

Reliable weather forecasting models play a crucial role in preparing for the harsh consequences of climate change by providing early warnings of extreme weather events for disaster mitigation. Numerous deep learning (DL)-based weather prediction models have been developed to forecast global weather under the supervision of the reanalysis mesh-gridded data \cite{Dueben2018,Weyn2019,Weyn2020,Rasp2020,Keisler2022}. FCN \cite{fourcastnet,Kurth2022} has shown exemplary performance in high-resolution data-driven global weather forecasting measured by the similarity of its forecasts against mesh-gridded reanalysis data, ERA5\cite{era5}. However, FCN's mesh-gridded forecasts could not be directly compared against the raw, sparse, non-uniform observations. Yet, the gold standard for weather and climate model performance is their ability to match observations. Further, because FCN is trained on reanalysis data, it is likely to have biases with respect to direct observations. To address these issues, we develop a DL-based \emph{grid-free} model that remaps and corrects the mesh-gridded forecasts to arbitrary locations in space and time. This methodology is sufficiently general that it can be applied to \emph{any} gridded forecast, not just from FCN.

Several data-driven approaches have been developed to refine mesh-gridded forecasts using sparse, non-uniform observational data \cite{Rasp2018,Amato2020,Zhang2022}. However, our objective is different from existing work in the following distinct ways: (1) The desired model should accept arbitrary output query locations, \emph{i.e.} it should be grid-free (existing works are only suitable for fixed locations); (2) Some observations may be missing randomly over space and time, so the desired model should handle random missing data seamlessly to fully utilize the available data at all locations and time steps; (3) The desired model should be capable of processing large spatio-temporal data from long forecasts, including all relevant variables, at high resolution, which can be highly memory-intensive.

To address these challenges, we develop a Fourier-based approach to learn continuous representations of the input data under the supervision of sparse, non-uniform observations. The model is grid-free since the continuous data can be reconstructed from the Fourier coefficients at any location. Moreover, because the network itself is independent of the output query, the model can naturally process missing data. To the best of our knowledge, this is the first time that a \emph{grid-free} DL model has been developed to re-map and correct mesh-gridded high-resolution weather forecasts to sparse, non-uniform observations.

\section{Methodology and Data}

\subsection{Type II Non-Uniform DFT}
We use Fourier coefficients for continuous representations of spatio-temporal data. Therefore, the projection of the continuous representation to spatial positions is the real part of type II non-uniform inverse discrete Fourier transform (NUIDFT), which can be written in matrix form as:
\begin{align}
    \cos(2\pi QM^T)F_{real}-\sin(2\pi QM^T)F_{img},
\end{align}
where $Q\in \mathbb{R}^{N\!\times\!2}$ denotes the query matrix, $M\in \mathbb{R}^{W\!H\!\times\!2}$ is the frequency basis, $F_{real}$ and $F_{img}\in \mathbb{R}^{W\!H\!\times\!1}$ denote the real and imaginary parts of the Fourier coefficients, respectively.

The sparse NUIDFT is developed to reduce the memory footprint by dividing the queries into different groups with respect to their actual length and truncating zero elements. For very high-resolution data, query locations can be randomly sub-sampled in each epoch to further reduce memory usage.

\subsection{DL Corrector-Remapper and Baselines}

\begin{figure*}[!t]
\centerline{\includegraphics[width=0.99\textwidth]{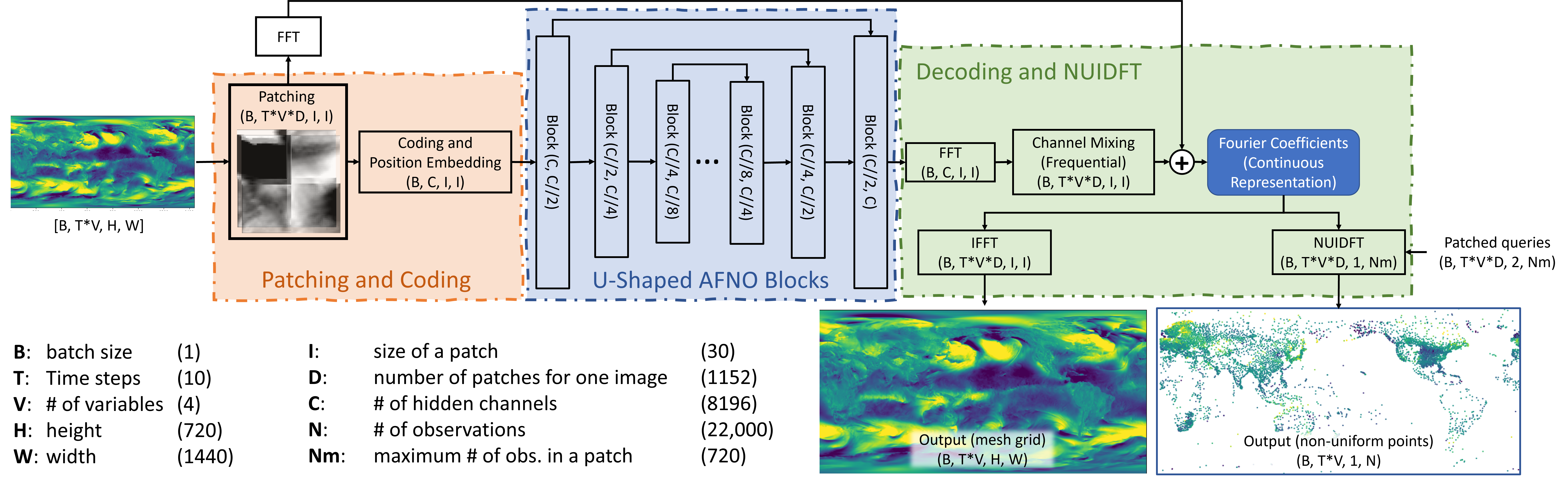}}
\caption{Flowchart of DL Corrector-Remapper (DLCR). DLCR consists of three parts: (i) image patching and coding; (ii) U-shaped AFNO blocks; and (iii) decoding and NUIDFT. The typical size of the output tensor is denoted under the name of each block. }
\label{fig1}
\end{figure*}

We develop DL-Corrector-Remapper (DLCR), a Fourier-based network that incorporates NUIDFT into the backbone of Adaptive Fourier Neural Operators (AFNO) \cite{afno}, to remap and bias-correct mesh-gridded forecasts to arbitrary locations in space and time, under the supervision of the gold standard -- sparse, non-uniform observational data. 

The overall structure of the proposed network is shown in Figure \ref{fig1}. Essentially, the output of the model is the Fourier coefficient, so one can use arbitrary queries to get arbitrary spatial outputs. Applying the inverse fast Fourier transform (IFFT) yields mesh-gridded output, while applying the NUIDFT with the query yields sparse observational output (see bottom right in Figure \ref{fig1}). 

In the \emph{patching process} 
, non-overlapping patches of the image are stacked in the channel dimension to form a token sequence \cite{afno}. Then, the channel coder compresses the channel size to reduce the memory footprint by accounting for the spatio-temporal and inter-variable dependencies. The \emph{U-shaped AFNO blocks} are the backbone of the proposed model. 
The channel dimension is halved for each subsequent block $k$ if $k<(K+1)//2$, and doubled if $k\geq(K+1)//2$, where $K$ denotes the total number of blocks. This U-shaped structure reduces the memory footprint without sacrificing the depth of the network. Moreover, \emph{skip connections} allow feature maps with more channels to flow in the network, so the narrowest block in the network does not reduce the complexity of the output.  

\begin{figure*}[!t]
\centerline{\includegraphics[width=0.90\textwidth]{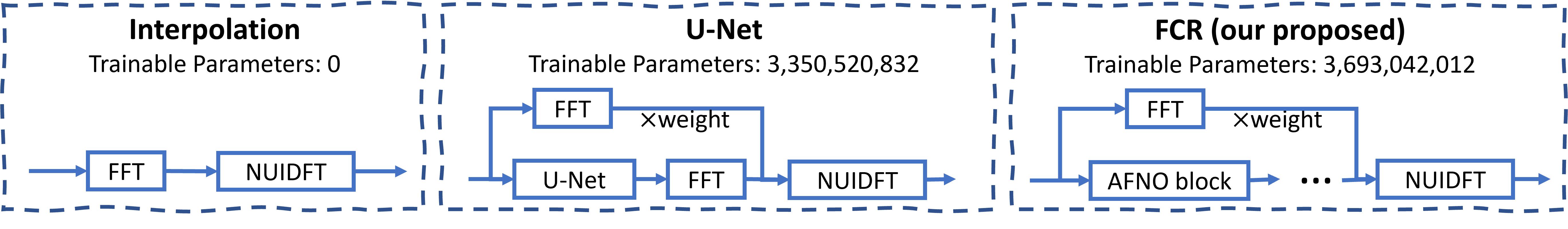}}
\caption{Model comparison of the interpolation, U-Net, and DLCR. These three methods utilize NUIDFT to reconstruct the spatio-temporal data from the continuous representation. The interpolation method has no trainable parameters, while U-Net and FCR both have $\sim$3 billion trainable parameters.}
\label{fig2}
\end{figure*}

The proposed network (DLCR) is compared to the NUIDFT-based interpolation and the widely used standard U-Net. Figure \ref{fig2} shows a schematic of these models and the corresponding number of trainable parameters.

\subsection{Loss function}

To correct biases of mesh-gridded forecasts whilst preserving their spatio-temporal structure, shapes, and patterns, we develop a novel loss function with two terms: (i) a magnitude difference term; and (ii) a structural similarity term,
\begin{align}
    \argmin_\theta ||&y_{sp}-\Gamma_\theta(x,q_{sp})||^2_2 + \lambda(1-LCC(y_{gd},\Gamma_\theta(x,q_{gd})))
\end{align}
where $y_{sp}$ denotes sparse, non-uniform observational data, $q_{sp}$ is the corresponding sparse output query, $y_{gd}$ denotes the reanalysis mesh-gridded data (ERA5), $q_{gd}$ is the mesh-gridded query, $\Gamma_\theta$ denotes the trainable model, $x$ denotes the input data, $\lambda$ denotes a scalar that controls the weight of the structural similarity term, and $LCC$ denotes the local cross-correlation \cite{Avants2008} (see Appendix A). 

\subsection{Datasets and Training}
\begin{figure}
\centerline{\includegraphics[width=0.99\textwidth]{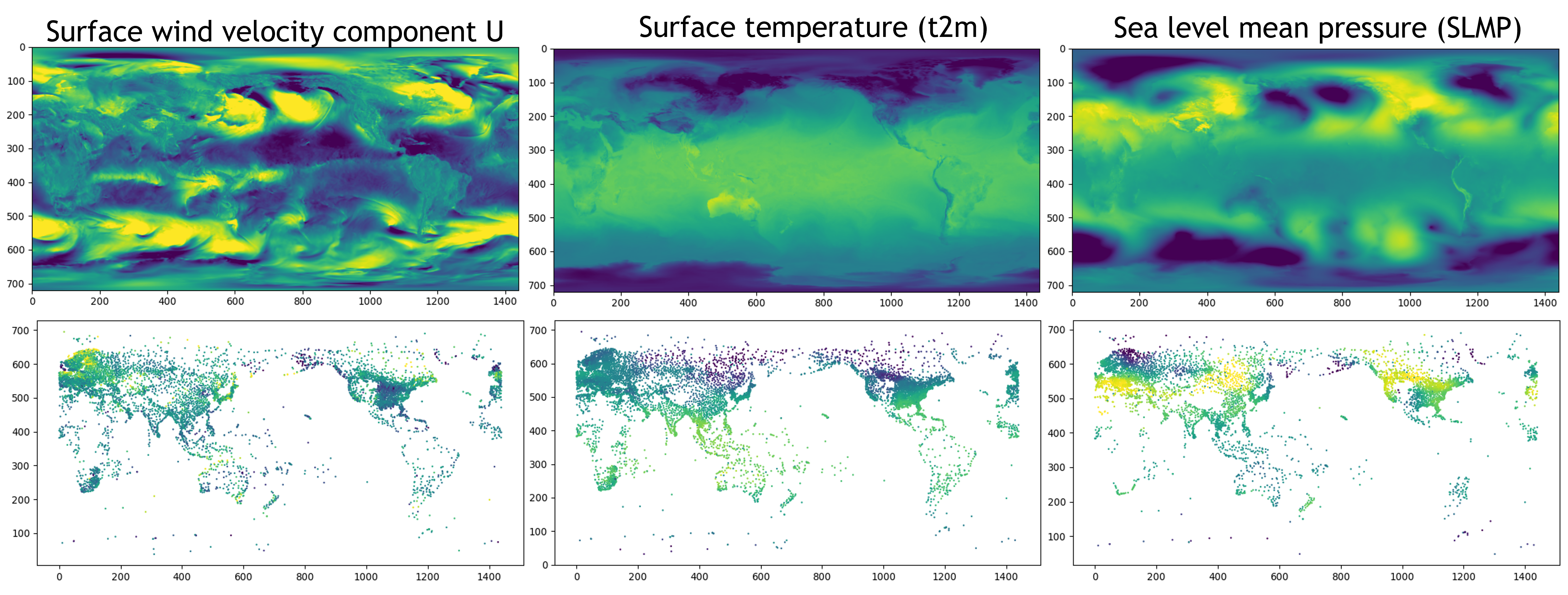}}
\caption{The mesh-gridded input (top) and non-uniform reference (bottom) data. Wind velocity V is omitted in this figure. The reference data is extremely sparse and missing randomly.}
\label{fig:data}
\end{figure}

The input dataset is FCN inference data from 2000 to 2018 at 0.25\degree resolution \cite{fourcastnet}. We use the global observational data acquired from National Centers for Environmental Information as the ground truth reference \cite{NOAA}. Four overlapping variables are selected: surface wind velocity components (U and V), surface temperature (t2m), and mean sea level pressure (MSLP). A data sample is shown in Figure \ref{fig:data}. The input and reference data are normalized by the global mean and variance of each variable.

Each forecast is five-days-long, with a 6-hourly time step, generated across 19 years, which yields 27360 samples in total. We use 23040 samples from 2000 to 2015 to train the model and 1440 samples from the year 2017 for validation. Moreover, 2\% of the observational sites are removed from the training set, so these positions are `unobserved' or \emph{hidden}. `Unobserved' time is used to test the reliability of future forecasts, while `unobserved' positions are used to test the model's ability to produce observation-quality forecasts for locations that it has never seen before. Since the reference data is extremely sparse, the output query is perturbed by Gaussian noise $\mathcal{N}(0,0.04)$ to address spatial over-fitting. 

The networks are trained on 64 A100 NVIDIA GPUs with 80 GB memory on the Selene supercomputer \cite{Selene}. Each training epoch takes 3000 seconds. The optimizer is ADAM, and the start learning rate is $10^{-3}$. We train the model for 10 epochs with a scheduler that reduces the learning rate by a factor of 0.7 every 3 epochs.

\section{Results}

\begin{figure*}[!t]
\centerline{\includegraphics[width=0.98\textwidth]{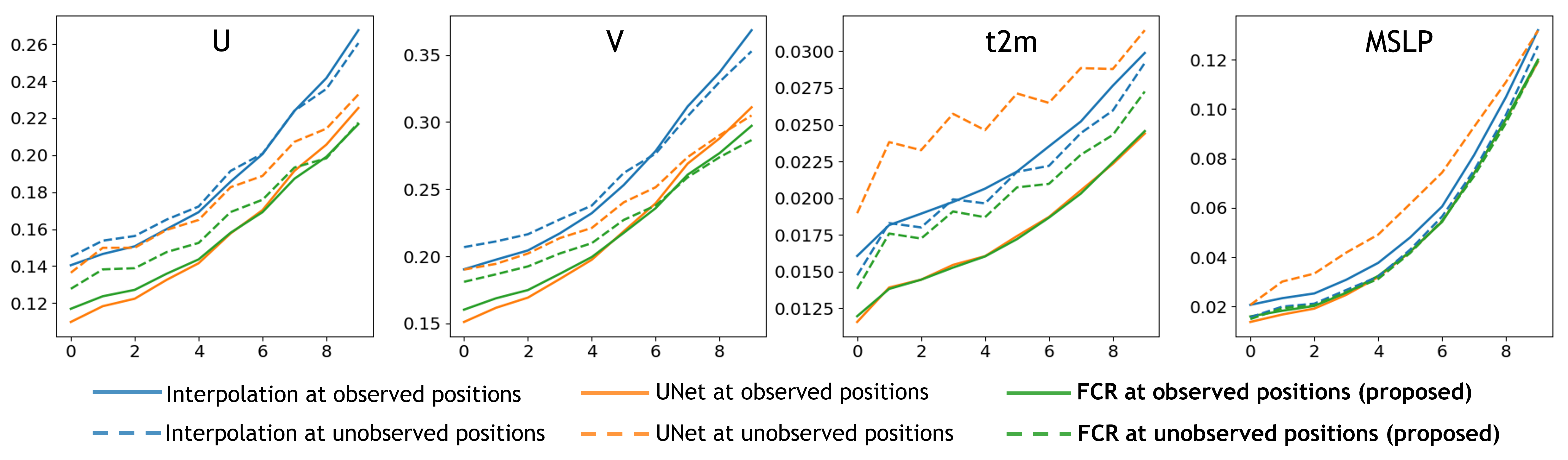}}
\caption{MSE of the output against the observation vs. prediction time step (lower is better). Our proposed model (DLCR) outperforms baselines across observed and unobserved locations. }
\label{fig:result}
\end{figure*}

Figure \ref{fig:result} shows the plots of mean square error (MSE) that are averaged over 80 instances across the year 2017 (validation). The horizontal axis denotes the time step, where each step corresponds to 12 hours. For observed positions, the performance of DLCR is close to the performance of U-Net, and they both outperform the interpolation baseline. For unobserved positions, DLCR outperforms the interpolation baseline and U-Net, and it performs better on more complicated variables (U and V), whereas the performance of the U-Net is even worse than the performance of the interpolation in estimating t2m and MSLP. Besides, the difference between the solid and dashed blue curves may indicate a small systematic bias in the unobserved locations. (see appendix for additional information) 

\section{Conclusion}

We propose a grid-free DL model (DLCR) to correct, remap, and fine-tune gridded uniform high-resolution global weather forecasts under the supervision of the gold standard ground truth -- observations -- which are sparse and non-uniform in space and time, can be noisy, and have gaps and missing data. DLCR outperforms baselines even when the output positions are unobserved. In this paper DLCR is developed for correcting and remapping FourCastNet's output, but the method itself can be applied to any forecast data. This work has implications for improving a broad range of data-driven and traditional NWP forecasting approaches by making their outputs comparable to and trainable on observational data. Importantly, DLCR allows arbitrary query locations and is trained to produce \emph{observation-quality} forecasts in `unobserved' locations, therefore it provides a way to improve the reliability of data-driven and NWP forecasts in locations on the globe and for time instances where observations do not exist.

\bibliographystyle{unsrt}
\bibliography{ref}

\appendix

\section{Appendix: Local Cross-Correlation}

Local cross-correlation (LCC) is a similarity metric that focuses on the similarity of patterns, structures, and high contrast boundaries between signals, written as 
\begin{align}
    LCC(A,B)=\frac{\Big(\sum_i\sum_j \big(A[i,j]-\Bar{A}[i,j]\big)\big(B[i,j]-\Bar{B}[i,j]\big)\Big)^2}{\sum_i \sum_j\Big(A[i,j]-\Bar{A}[i,j]\Big)^2 \sum_i\sum_j \Big(B[i,j]-\Bar{B}[i,j]\Big)^2},
\end{align}
where $A$ and $B$ denote two arbitrary 2-dimensional images, and $\Bar{A}$ denotes the local mean that is calculated by the convolution of the image $A$ and an averaging kernel $V$, \emph{i.e.}, 
\begin{align}
    \Bar{A}[i,j]=\sum_n \sum_m A[i-n,j-m]V[n,m].
\end{align}
In this work, V is chosen to be a $5\times5$ uniform matrix whose sum is normalized to 1. The size of the kernel controls the scale of the detectability. LCC with a large kernel only captures large structures and ignores subtle changes, while LCC with a small kernel captures subtle variations yet is more vulnerable to noise.

\section{Appendix: Additional Figures}
\begin{figure*}[!t]
\centerline{\includegraphics[width=0.9\textwidth]{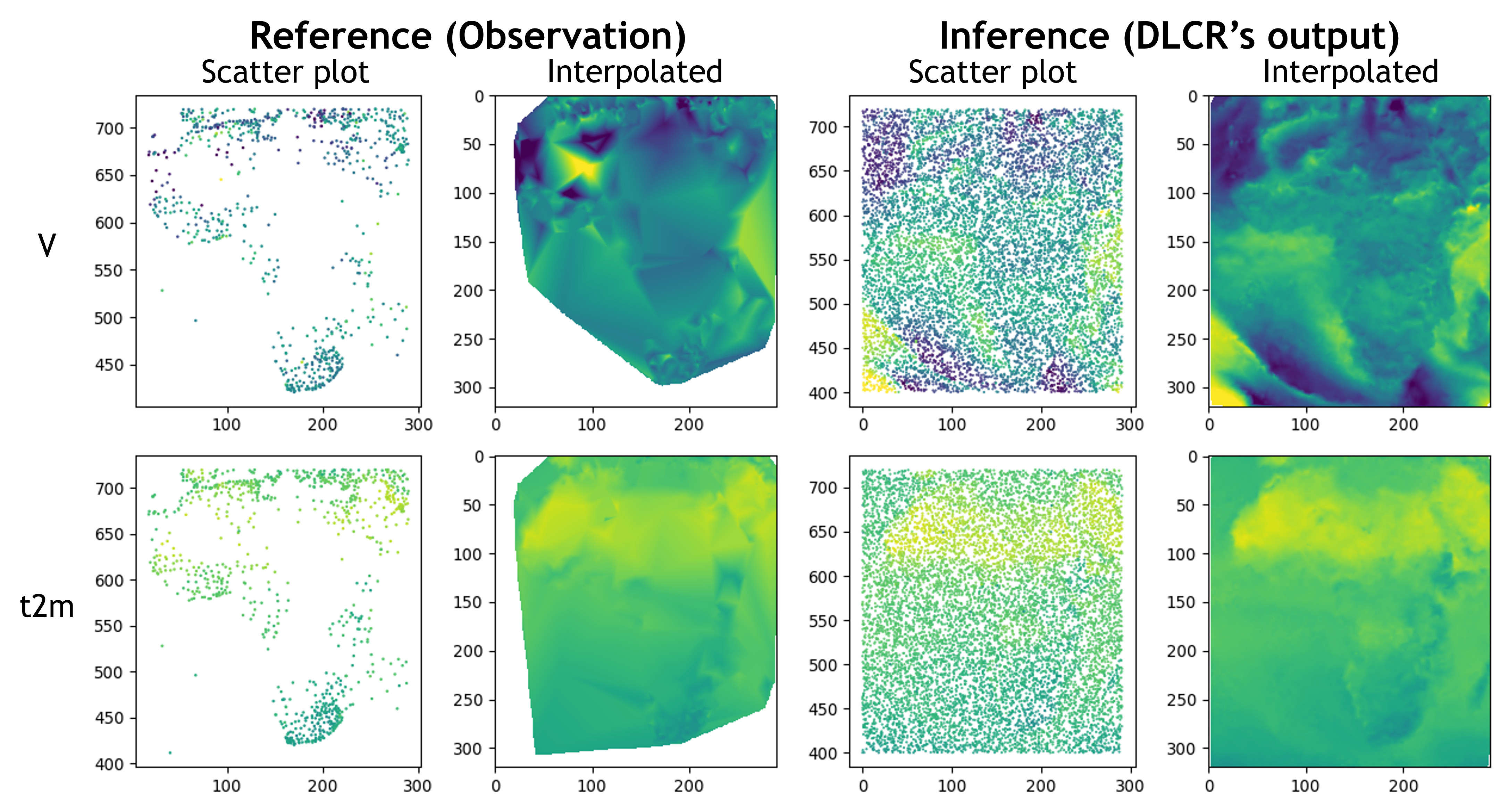}}
\caption{Figures of the reference and DLCR inference output for the region of Africa. The DLCR inference output is generated via a random query unseen by the model. It is clear that interpolation of DLCR's output, after querying DLCR densely over the African continent, has significantly better qualitative performance in regions of Africa with very sparse observations, whereas interpolation of the raw observations produces unphysical results in these regions. Thus the benefit of a grid-free model that can be queried at arbitrary locations is evident.}
\label{fig:6}
\end{figure*}

Figure \ref{fig:6} shows the scatter plot and the interpolated mesh-gridded image of the reference observation data, as well as the scatter plot and the interpolated output inference with a random query (6000 locations). The display region covers -40\degree to 40\degree of latitude and -22.5\degree to 50\degree of longitude, which corresponds to the continent of Africa. These query positions are not necessary to be on the mesh grid or observed. The randomly queried output of the proposed method still looks realistic in the region with a limited number of observational sites. 

\begin{figure*}[!t]
\centerline{\includegraphics[width=0.99\textwidth]{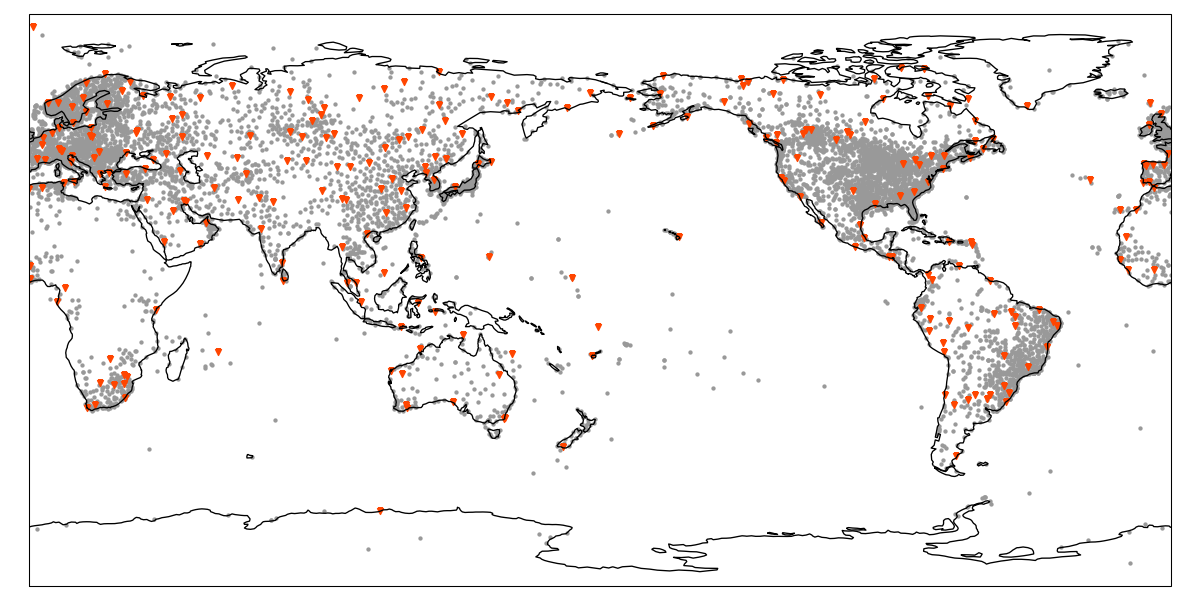}}
\caption{Figure showing the locations of the surface stations. Observed positions are indicated by grey dots, and unobserved positions are indicated by orange pins.}
\label{fig:7}
\end{figure*}

Figure \ref{fig:7} shows the observation positions in this work, where hidden positions are labelled as orange triangles. The world map is evenly divided into 48 patches, and these hidden positions are randomly selected in each patch.

\end{document}